\documentclass[a4paper,11pt]{article}
\usepackage{pos}

\newcommand{\MSb}{{\overline{\rm MS}}}

\title{{Generalized parton distributions of the proton from lattice~QCD}}
\ShortTitle{GPDs of the proton from lattice QCD}

\author[a,b]{Constantia Alexandrou}
\author[c]{Krzysztof Cichy}
\author[d]{Martha Constantinou}
\author[a,b]{Kyriakos Hadjiyiannakou}
\author[e]{Karl Jansen}
\author*[d]{Aurora Scapellato}
\author[f]{Fernanda Steffens}

\affiliation[a]{Department of Physics,  Temple University,  Philadelphia,  PA 19122 - 1801, USA}

\affiliation[b]{Department of Physics, University of Cyprus,  P.O. Box 20537,  1678 Nicosia, Cyprus}
\affiliation[c]{The Cyprus Institute, 20 Kavafi Str., Nicosia 2121, Cyprus}
\affiliation[d]{Faculty of Physics, Adam Mickiewicz University, Uniwersytetu Pozna\'nskiego 2, 61-614 Pozna\'{n}, Poland}
\affiliation[e]{Deutsches Elektronen-Synchrotron DESY, Platanenallee 6, 15738 Zeuthen, 
Germany}
\affiliation[f]{Institut f\"ur Strahlen- und Kernphysik, Rheinische Friedrich-Wilhelms-Universit\"at Bonn, Nussallee 14-16, 53115 Bonn}

\emailAdd{alexand@ucy.ac.cy}  \emailAdd{krzysztof.cichy@gmail.com} \emailAdd{marthac@temple.edu}\emailAdd{k.hadjiyiannakou@cyi.ac.cy}\emailAdd{karl.jansen@desy.de}\emailAdd{aurora.scapellato@temple.edu} \emailAdd{fsteffens@uol.com.br}

\abstract{Generalized parton distributions (GPDs) are among the most fundamental quantities for describing
the internal structure of hadrons. In this work, we present results on isovector GPDs of the proton obtained within lattice Quantum Chromodynamics. We
use the quasi-distribution formalism and perform the calculations on an ensemble of $N_f = 2 + 1 + 1$ twisted mass fermions, with pion mass $M_\pi=260$~MeV and lattice spacing $a\simeq 0.093$~fm. Results are presented for
unpolarized, helicity, and transversity GPDs at zero and nonzero skewness with controlled statistical uncertainties. Comparisons with their forward limit show qualitative features anticipated from
model calculations.}

\FullConference{%
 The 38th International Symposium on Lattice Field Theory, LATTICE2021
  26th-30th July, 2021
  Zoom/Gather@Massachusetts Institute of Technology
}

\usepackage{slashed}
\begin{document}
\maketitle

\section{Introduction}
One of the open questions in Quantum Chromodynamics (QCD) is how quarks and gluons (collectively called \textit{partons}) give rise to the global properties of hadrons and nuclei, such as the mass and spin. In this context, generalized parton distributions (GPDs) are among the most suitable candidates to map the 3D structure of hadrons. Introduced in the '90s as generalization of parton distribution functions (PDFs) and elastic form factors~\cite{Ji:1996ek,Radyushkin:1996nd,Mueller:1998fv}, GPDs give also access to generalized form factors, which encode e.g.\ the energy-momentum tensor. Therefore, through appropriate sum rules and Fourier transforms, GPDs provide insight into the orbital angular momentum of partons~\cite{Ji:1996ek}, the pressure and shear forces inside hadrons~\cite{Polyakov:2002yz}, as well as  the correlation between the longitudinal momentum of partons and their position in the transverse spatial plane in a hadron~\cite{Burkardt:2002hr}. Despite  their importance, very little is known about GPDs so far and it is expected that the planned experiments at the 12 GeV upgrade program at Jefferson Lab~\cite{Dudek:2012vr,Burkert:2018nvj} and at the Electron-Ion Collider (EIC)~\cite{AbdulKhalek:2021gbh} will be essential to pin down these quantities. On the other hand, lattice QCD may also be a complementary framework to extract GPDs directly from the QCD Lagrangian.

Formally, as any other partonic distribution, GPDs are defined in terms of quark bilinear operators sitting on the light front~\cite{Collins:2011zzd}. This makes their extraction from lattice QCD highly non-trivial, because light-cone separations are not accessible in Euclidean spacetime using a finite lattice spacing. To overcome this limitation, different approaches relying to some extent on the factorization framework have been developed, namely quasi-distributions~\cite{Ji:2013dva}, pseudo-distributions~\cite{Radyushkin:2017cyf}, ``good'' lattice cross sections~\cite{Ma:2014jla,Ma:2014jga,Ma:2017pxb}, ``OPE without OPE''~\cite{Chambers:2017dov} and auxiliary quark methods~\cite{Braun:2007wv,Detmold:2005gg} (see e.g.~\cite{Cichy:2018mum,Ji:2020ect,Constantinou:2020pek,Cichy:2021lih} for reviews in the field) and very promising results from lattice QCD have been obtained over the last few years. These involve especially the computation of quark and gluon PDFs in hadrons with different spin.
 
 In this work, we extend the methodologies applied to PDFs~\cite{Alexandrou:2018pbm,Alexandrou:2018eet,Alexandrou:2019lfo} and compute the isovector GPDs of the proton with different polarizations, using the quasi-distribution approach~\cite{Ji:2013dva}. In particular, we compute four chiral-even GPDs, $H,E,\tilde{H},\tilde{E}$ (two for spin-averaged and two for longitudinal polarization) and four chiral-odd GPDs, $H_T, E_T, \tilde{H}_T, \tilde{E}_T$ (for a transversely polarized proton). All the GPDs are extracted as functions of the three kinematic variables they depend on: $x$ - the longitudinal momentum fraction of the struck quark with respect to the momentum of the target, $\xi$ - known as \textit{skewness}, that is half the change in the momentum fraction of the struck parton with respect to the target and $t$ - the squared four-momentum transferred to the target. For a complete description of our lattice results we refer the Reader to~\cite{Alexandrou:2020zbe,Alexandrou:2021bbo}.

\section{GPDs from quasi-distributions}
Within the quasi-distribution approach, GPDs are accessed through purely spatial correlation functions with boosted hadrons, that have different momenta in the initial and final states. The starting point is the matrix element
 \begin{equation}
h[\Gamma,z,P_f,P_i,\mu_0]={Z}_\Gamma(z,\mu_0)\langle N(P_f)|O_\Gamma|N(P_i)\rangle\,\,\mbox{with }\quad O_\Gamma=\bar{\psi}(0)\Gamma W(0,z)\psi(z)\,,
\label{eq:ME}
 \end{equation}
where $|N(P_i)\rangle$ and $|N(P_f)\rangle$ are the initial (source) and final (sink) state of the proton with four-momentum $P_i$ and $P_f$ respectively, $\psi$ is the doublet of light quarks and $W$ is a straight Wilson line of length $z$. The Dirac matrix $\Gamma$ specifies the type of GPD: $\gamma_\mu$ for unpolarized, $\gamma_\mu\gamma_5$ for helicity and $\sigma^{\mu\nu}$ for transversity GPDs. Each matrix element, $h$, is renormalized in a given scheme at a scale $\mu_0$, through renormalization functions $Z_{\Gamma}(z,\mu_0)$. The renormalized matrix elements of Eq.(\ref{eq:ME}) are related to the GPDs by continuum decompositions (see e.g.~\cite{Ji:1998pc}). For example, at the leading twist in Euclidean space and for the unpolarized case, they take the form
\begin{equation}
\langle N(P_f)|O_{\gamma_\mu}|N(P_i)\rangle=\bar{u}_N(P_f)\left[ \gamma_\mu F_H(z,P_3,t,\xi)-i\frac{\sigma_{\mu\nu}}{2m_N}Q_\mu F_E(z,P_3,t,\xi)\right] u_N(P_i)\,,
\label{eq:decomposition}
\end{equation}
where $m_N$ is the nucleon mass, $Q=P_f-P_i$ and $P_3$ is the average momentum boost, $P=(P_i+P_f)/2$, along the $z$-direction. Analogous decompositions exist for the helicity and transversity GPDs. In Eq.(\ref{eq:decomposition}) $F_H$ and $F_E$ play the role of \textit{form factors} of the GPDs, because they reduce  to the physical GPDs upon Fourier transform to momentum space
\begin{equation}
\widetilde{G}(x,\xi,t,\mu_0,P_3)=\int_{-\infty}^{+\infty}dz\, e^{-iP_3xz}\,F_G(z,\xi,t,P_3,\mu_0)\,,\quad  G=H,E,\ldots  
\end{equation}
and matching within Large Momentum Effective Theory (LaMET)~\cite{Ma:2014jla,Ji:2020ect},
\begin{equation}
\label{eq:matching}
\resizebox{.95\hsize}{!}{$
\widetilde{G}(x,\xi,t,\mu_0,(\mu_0)_3,P_3) = \int_{-1}^1 \frac{dy}{|y|}\, C_G \left(\frac{x}{y},\frac{\xi}{y},\frac{\mu}{y P_3},\frac{(\mu_0)_3}{y P_3},r\right) G(y,t,\xi,\mu)+\mathcal{O}\left(\frac{m^2}{P_3^2},\frac{t}{P_3^2},\frac{\Lambda_{\rm QCD}^2}{x^2P_3^2}\right)\,.$}
\end{equation}
Here, $C_G$ is the matching kernel, known at present to 1-loop level in perturbation theory. In this work we use the expression derived in~\cite{Liu:2019urm}, which brings the quasi-GPDs $\widetilde{G}$ in the RI-MOM scheme, at a scale $\mu_0$ ($r=\mu_0^2/(\mu_0)_3^2$), to the $\MSb$ scheme GPDs at scale of reference $\mu$. Quasi- and light-cone GPDs differ by power corrections in $1/P_3^2$ and therefore, the nucleon should be boosted with a very large momentum to improve convergence. However, in lattice computations, the maximal achievable boost remains limited by an exponential increase of the noise-to-signal ratio and by cutoff effects, that are enhanced if $P_3$ becomes comparable to the inverse of the lattice spacing (see Table~1 of~\cite{Constantinou:2020pek} for typical parameters in lattice QCD calculations).

\section{Lattice setup}
We use an ensemble of $N_f=2+1+1$ maximally twisted mass fermions~\cite{ExtendedTwistedMass:2021gbo} with volume $V=32^3\times 64$, pion mass $M_\pi\simeq 260$~MeV, $M_\pi L\simeq 4$ and lattice spacing $a\simeq 0.093$~fm. On this ensemble we compute the matrix elements in Eq.(\ref{eq:ME}) and isolate the isovector combination $u-d$, by using the Pauli matrix $\tau_3$. The matrix elements are obtained by the following ratio of appropriate two-point functions, $C_{2}$, and three-point functions $C_{3}$ 
\begin{equation}
\resizebox{.95\hsize}{!}{$
R(\mathcal{P}_\mu,\vec{P}_f,\vec{P}_i, t_{s}; t_{ins})=\frac{C_{3}(\mathcal{P}_\mu,\vec{P}_f,\vec{P}_i, t; t_{ins})}{C_{2}(\mathcal{P}_0;\vec{P}_f,t_{s})}\sqrt{\frac{C_{2}(\mathcal{P}_0;\vec{P}_i,t_s-t_{ins}) C_{2}(\mathcal{P}_0;\vec{P}_f,t_{ins}) C_{2}(\mathcal{P}_0;\vec{P}_f,t_{s})}{C_{2}(\mathcal{P}_0;\vec{P}_f,t_s-t_{ins}) C_{2}(\mathcal{P}_0;\vec{P}_i,t_{ins}) C_{2}(\mathcal{P}_0;\vec{P}_i,t_{s})}}$}\,,
\label{eq:ratio}
\end{equation}
that for sufficiently large time-separations between the sink and the insertion ($t_s-t_{ins}\gg 1$, with $t_{ins}\gg 1$) provides an estimate of the matrix element~(\ref{eq:ME}) on the ground state. In Eq.(\ref{eq:ratio}), $\mathcal{P}_\mu$ are parity projection matrices ($\mathcal{P}_0=\frac{1+\gamma_0}{2}$, $\mathcal{P}_i=\frac{1+\gamma_0}{2}\gamma_5\gamma_i$) entering through the three-point functions. 
The latter are computed using sequential inversions through the sink, with $t_s=12a\simeq 1.13$~fm to control excited states effects~\cite{Alexandrou:2019lfo}. The signal of each correlator in~(\ref{eq:ratio}) is optimized using momentum smeared interpolating fields~\cite{Bali:2016lva} and in the smearing kernel we use a global phase, $e^{i\vec{\eta}\vec{P}}$, that is kept parallel to the initial and final nucleon boosts, in order to avoid potential issues related to rotational symmetry breaking. Doing so, quark propagators cannot be reused for different kinematic setups $(P_3,\xi,t)$, but, on the other hand, we can optimize the signal for each correlator by tuning the free parameter $\vec{\eta}$. In addition, the appropriate combinations of the projectors, $\mathcal{P_\mu}$, and the Dirac matrix, $\Gamma$, depend on the frame and on the kinematic setup~\cite{Alexandrou:2020zbe,Alexandrou:2021bbo}, as specified in the following. Throughout this work we adopt the Breit frame (where GPDs are typically defined) and hence, the initial and final momenta are symmetric with respect to the average momentum boost, here taken to be nonzero only along the $z$-direction. Therefore, $\vec{P}_f=P_3\hat{z}+\vec{Q}/2$ and $\vec{P}_i=P_3\hat{z}-\vec{Q}/2$, with $Q^2=-t$. As a consequence, we choose the momentum transferred so that $\vec{Q}/2$ can be expressed in terms of a Fourier momentum $2\pi \vec{n}_q/L$, being $n_q$ integer and $L$ the spatial extent of the lattice.

To test the feasibility of the whole approach, we extract all GPDs at zero and nonzero skewness $\xi$, where $\xi=-Q_3/(2P_3)$. At $\xi=0$, we study the momentum dependence of the GPDs and, at the intermediate boost, we test the effect of a nonzero $\xi$. In Table~\ref{tab:stat_GPDs}, we summarize the parameters of the calculation and in Table~\ref{tab:stat_PDFs}, we report the statistics used for unpolarized, helicity and transversity PDFs that are extracted on the same gauge ensemble for a qualitative comparison with the GPDs. 
\begin{table}[h!]
\hspace*{0.5cm}
\resizebox{0.4\textwidth}{!}{
\begin{minipage}{0.49\textwidth}
\begin{center}
\renewcommand{\arraystretch}{1.4}
\begin{tabular}{|cccc|cc|}
\hline
$P_3$ [GeV] & $\quad \vec{Q}$ $[\frac{2\pi}{L}]\quad$ & $-t$ [GeV$^2$] & $\xi$ & $N_{\rm confs}$ & $N_{\rm meas}$\\
\hline
0.83 &(0,2,0)  &0.69  &0      & 519  & 4152\\
1.25 &(0,2,0)  &0.69  &0      & 1315  & 42080\\
1.67 &(0,2,0)  &0.69  &0      & 1753  & 112192\\
1.25 &(0,2,2)  &1.02  & 1/3   & 417  & 40032\\
1.25 &(0,2,-2) &1.02  & -1/3  & 417  & 40032 \\
\hline
\end{tabular}
\begin{minipage}{11cm}
\caption{Statistics for the GPDs, at each $P_3$, $\vec{Q}$ and $\xi$.}
\label{tab:stat_GPDs}
\end{minipage}
\end{center}
\end{minipage}}
\hspace*{2.1cm} 
\resizebox{0.42\textwidth}{!}{
\begin{minipage}{0.49\textwidth}
\hspace*{1.5cm} \begin{tabular}{|c|cc|}
\hline
$P_3$ [GeV] & $N_{\rm confs}$ & $N_{\rm meas}$\\
\hline
0.83 & 194 & 1560\\
\hline 
1.25  &  731  & 11696\\
\hline 
1.67  & 1644  & 105216\\
\hline
\end{tabular}
\caption{\hspace{0.2cm}Statistics for the PDFs.}
\label{tab:stat_PDFs}
\end{minipage}}
\end{table}

\begin{figure}[h!]
    \hspace{-0.3cm}
    \includegraphics[scale=0.5]{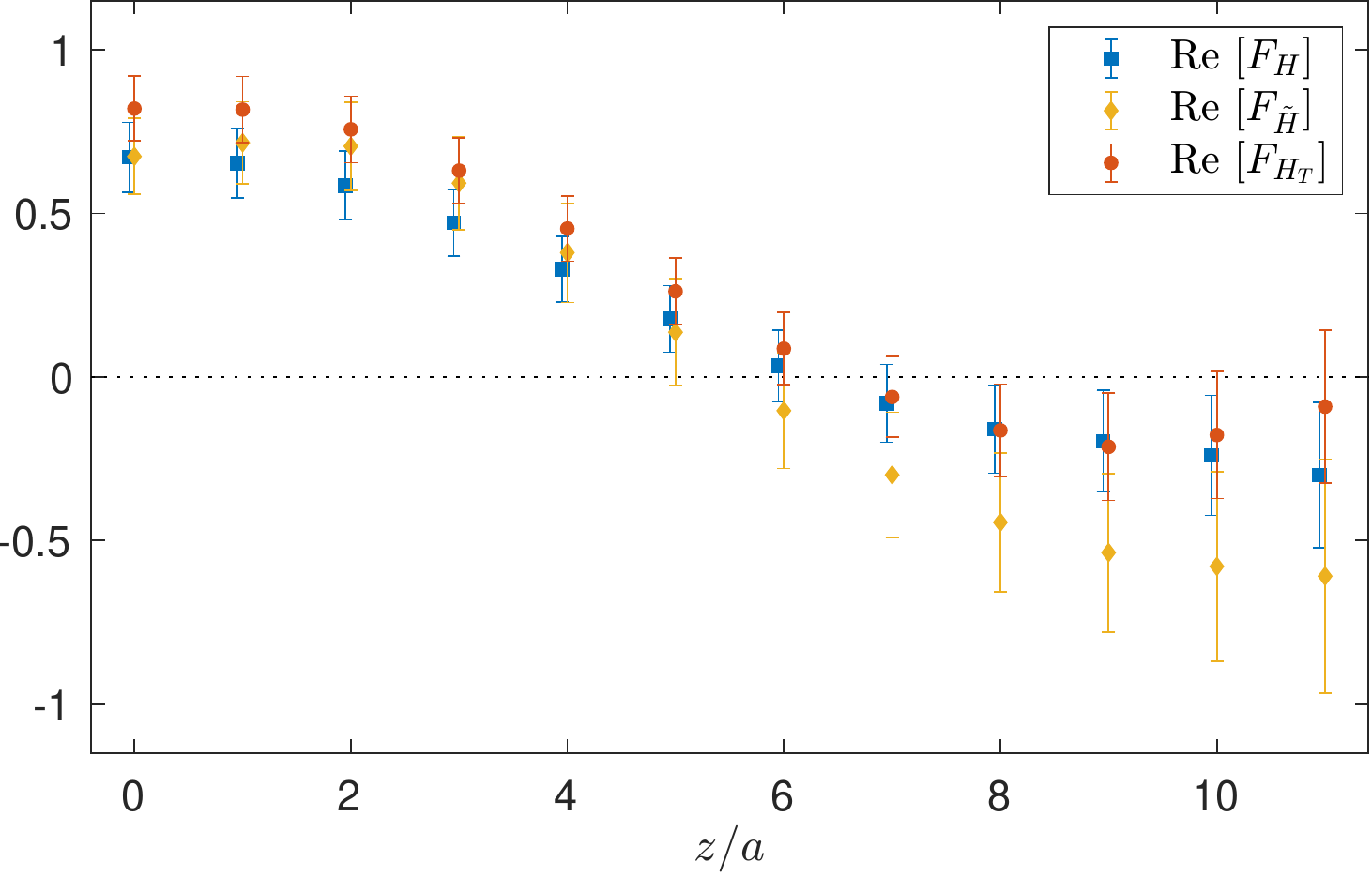}
    \includegraphics[scale=0.5]{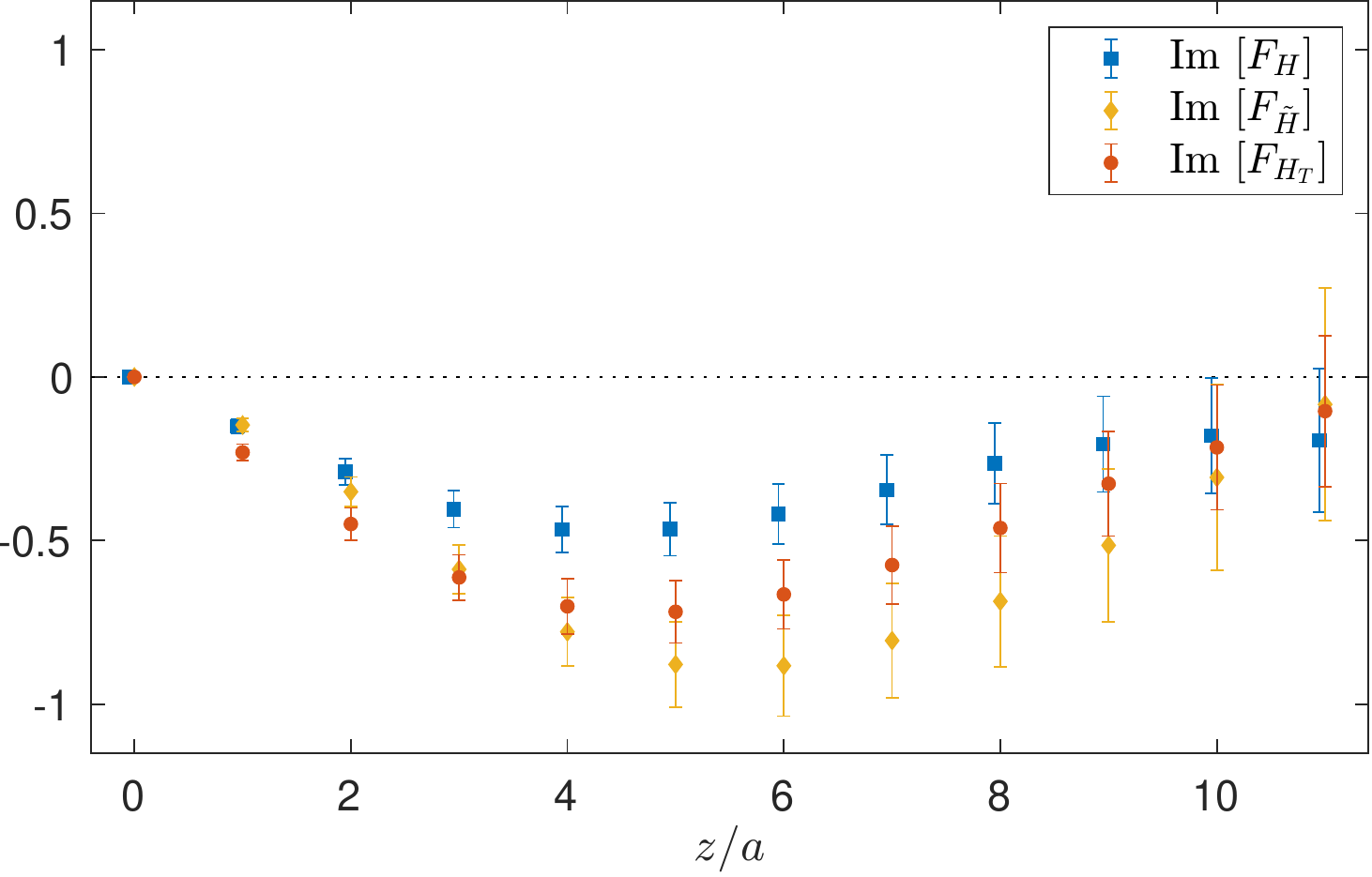}
    
    \includegraphics[scale=0.5]{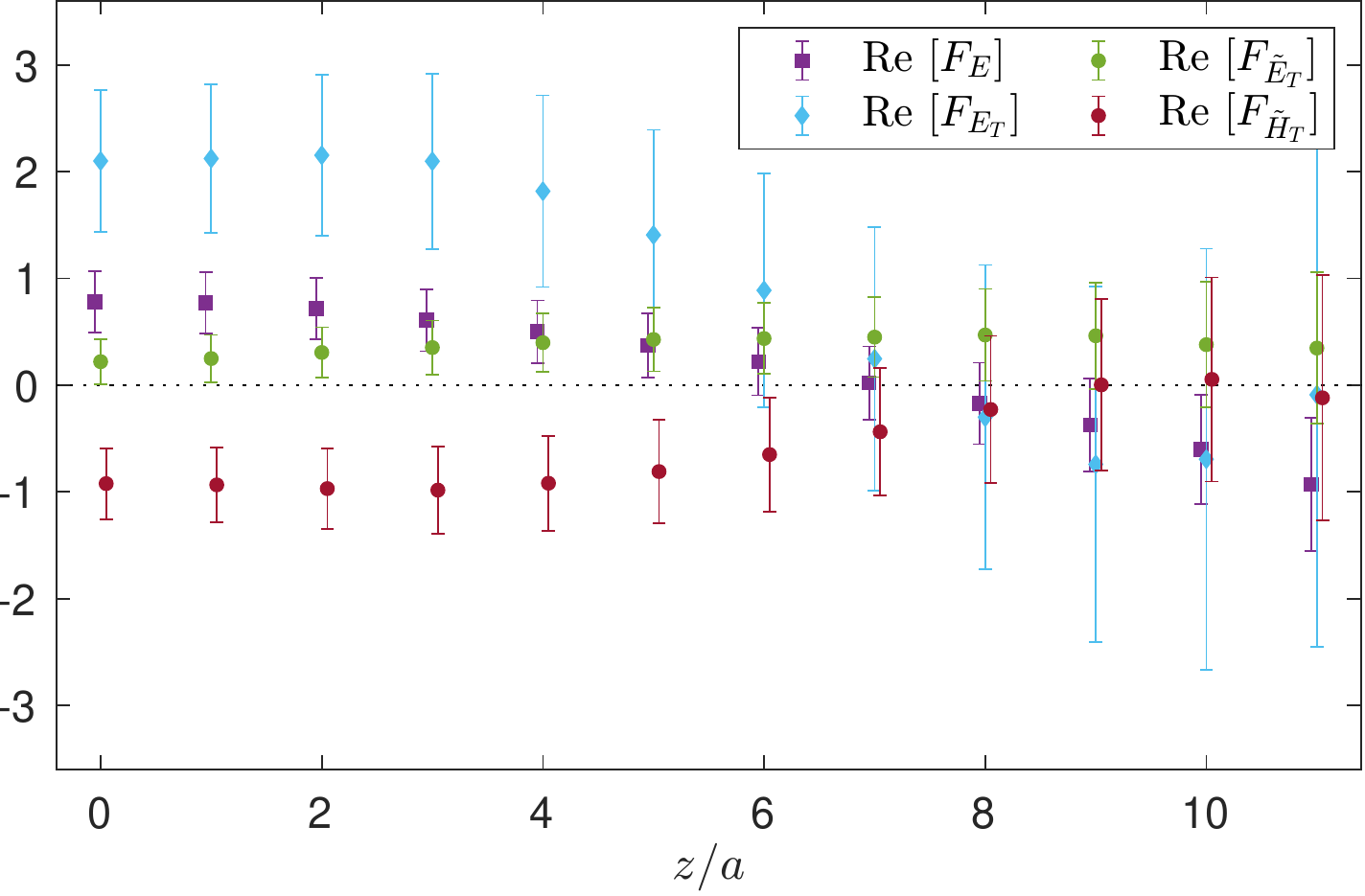}\hspace{0.3cm}
    \includegraphics[scale=0.5]{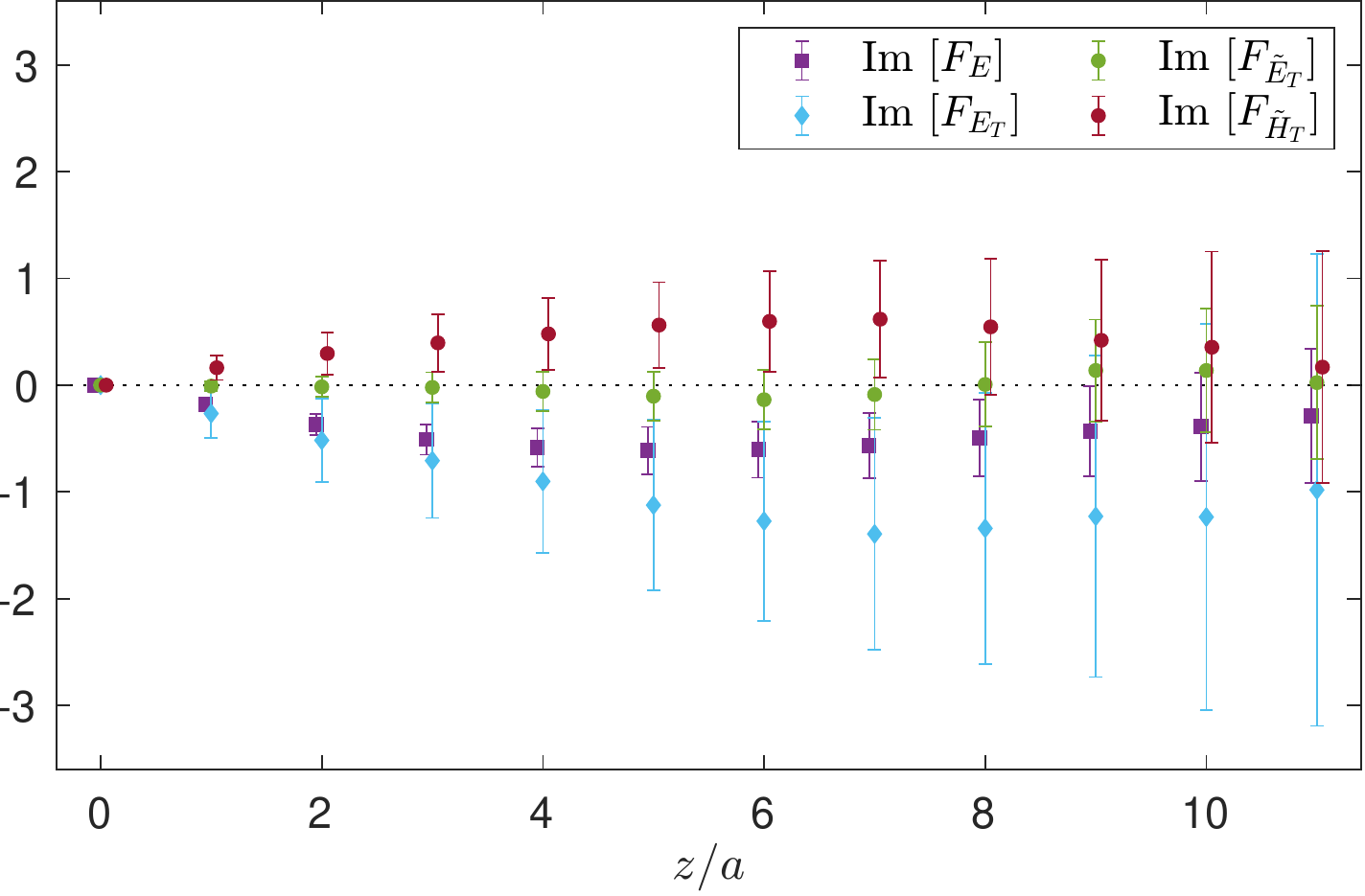}
    \caption{Real (left) and imaginary (part) part of the matrix elements for the leading (top) and subleading GPDs (bottom) at $\lbrace P_3=1.67,\xi=0,-t=0.69$GeV$^2\rbrace$. Data are renormalized in a variant of the RI-MOM scheme at $(a\mu_0)^2\approx 2.57$.}
    \label{fig:FHs}
\end{figure}

With these choices of momenta, the Euclidean decompositions in Eq.~(\ref{eq:decomposition}) require two matrix elements, with projectors $\mathcal{P}_0$, $\mathcal{P}_1$ and $\Gamma=\gamma_0$, to disentangle $F_H$ and $F_E$. For the helicity case, however, the kinematic factor of $\tilde{E}$ vanishes at $\xi=0$ and so, at $\xi=0$, we can only extract $\tilde{H}$ using the projector $\mathcal{P}_3$ and $\Gamma=\gamma_5\gamma_3$; at nonzero $\xi$ we need an additional matrix element, from $\mathcal{P}_2$, to disentangle $\tilde{H}$ from $\tilde{E}$. These matrix elements are related to the form factors of the GPDs via kinematic relations (dependent on $P_3$, $P_i$, $P_f$), that  can be found in~\cite{Alexandrou:2020zbe} for the unpolarized and helicity GPDs. The transversity GPDs, instead, can receive contributions from up to four independent matrix elements with $\Gamma=\sigma^{3j}$ ($j=1,2$) and therefore they are the most challenging GPDs to extract. With our setup, the only exceptions are $F_{H_T}$ and $F_{\tilde{E}_T}$ at $\xi=0$, that can be obtained from only one matrix element, namely for $\mathcal{P}_2$, $\sigma_{31}$ and $\mathcal{P}_3$, $\sigma_{31}$, respectively. We refer to~\cite{Alexandrou:2021bbo} for the specific expressions and decompositions. As an example, in Fig.~\ref{fig:FHs}, we summarize the results of the renormalized form factors at $\lbrace \xi=0$, $-t=0.69$~GeV$^2\rbrace$ and  $P_3=1.67$~GeV. 
The renormalization scheme matches the one of the matching kernel $C_G$ in Eq.~(\ref{eq:matching}) and, specifically, it is a variant of the RI-MOM scheme, namely $\slashed{p}$-projection for unpolarized and helicity and minimal-projection for transversity GPDs. As we can see from Fig.~\ref{fig:FHs}, the form factors of the leading GPDs (top) have similar uncertainties and the real value at $z=0$ is always positive. In fact, $F_H$, $F_{\tilde{H}}$ and $F_{\tilde{H}_T}$ reduce to the well-known form factors $F^{u-d}_1(Q^2)$, $g^{u-d}_A(Q^2)$ and $g^{u-d}_T(Q^2)$. The subleading contributions have instead large uncertainties, especially $F_{E_T}$ and $F_{\tilde{H}_T}$, because of their complex kinematics. We also find that $F_{\tilde{E}_T}$, expected to be odd under $\xi\rightarrow -\xi$~\cite{Meissner:2007rx,Diehl:2003ny}, is compatible with zero at all $z$-values. Similar qualitative conclusions also hold for the smaller momenta considered in this work.

Once the form factors have been computed, we perform the Fourier transform in the $z$-space using the Backus-Gilbert method~\cite{BackusGilbert,Karpie:2018zaz}, whose implementation for quasi-GPDs is described in detail in~\cite{Alexandrou:2021bbo,Alexandrou:2020zbe}. Finally, from quasi-GPDs we get the physical GPDs by inverting the matching formula~(\ref{eq:matching}).
\section{Momentum dependence of the GPDs}
\begin{figure}
    \includegraphics[scale=0.5]{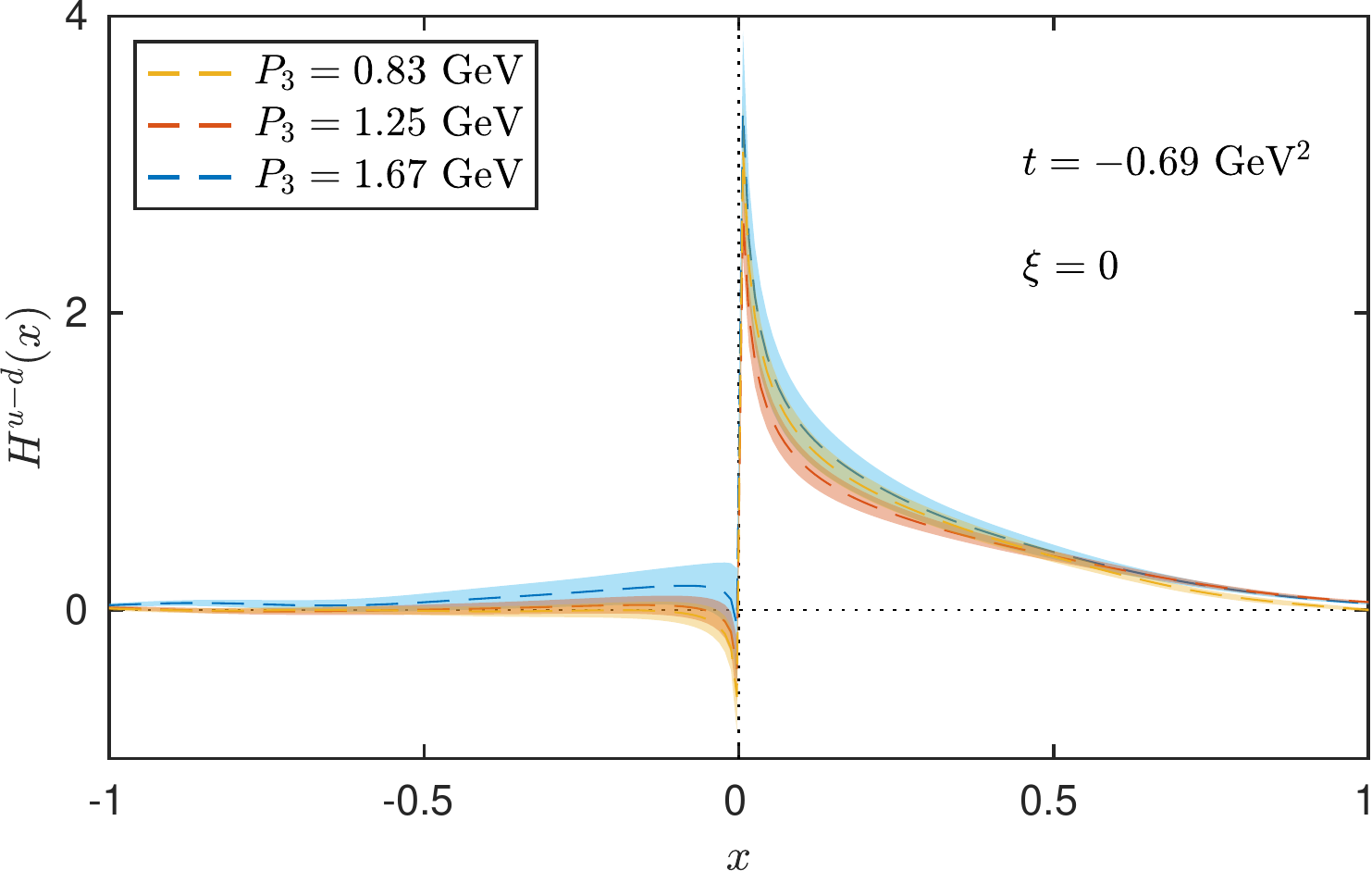}
    \includegraphics[scale=0.5]{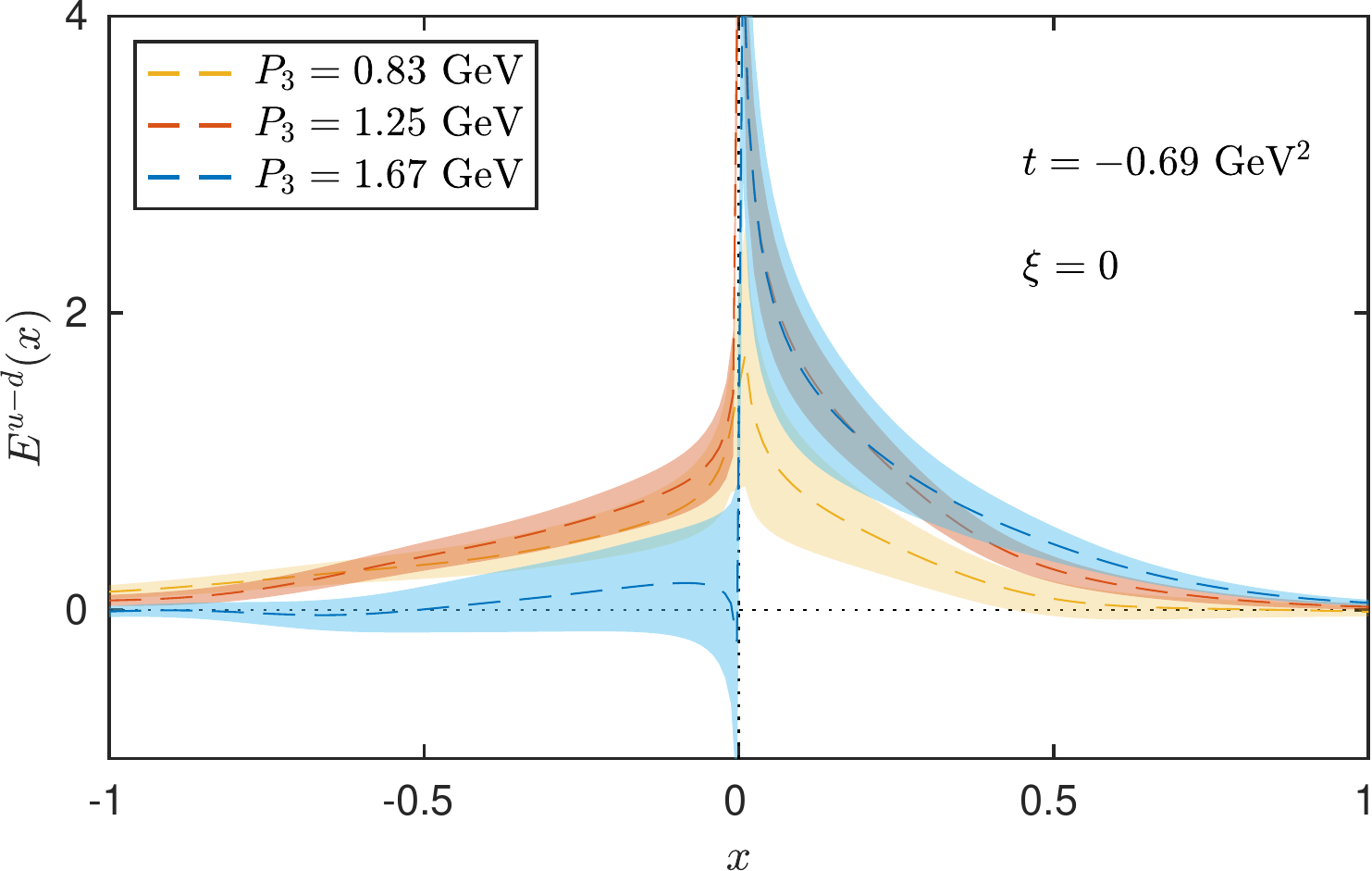}
    
    \includegraphics[scale=0.5]{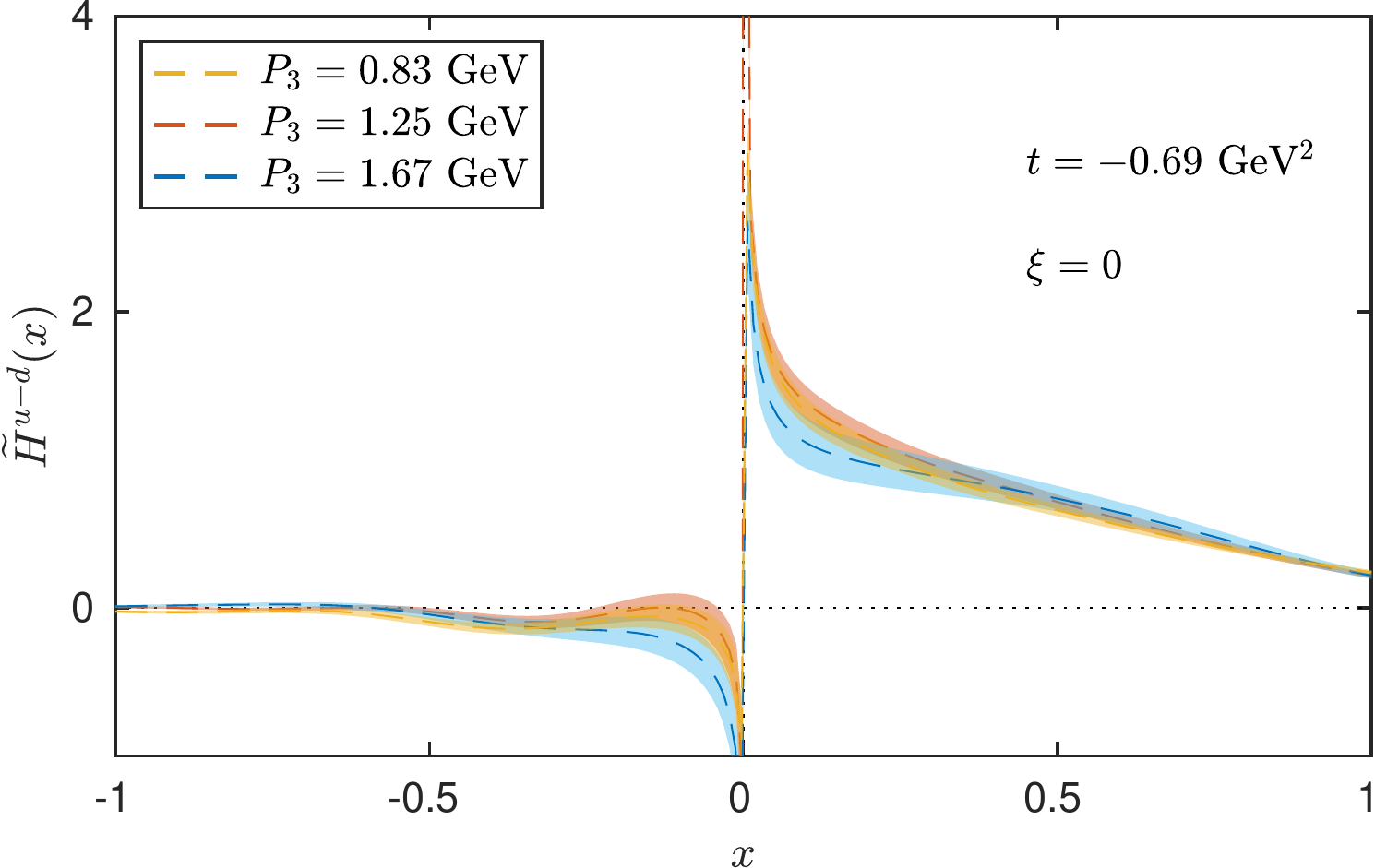}
    \includegraphics[scale=0.5]{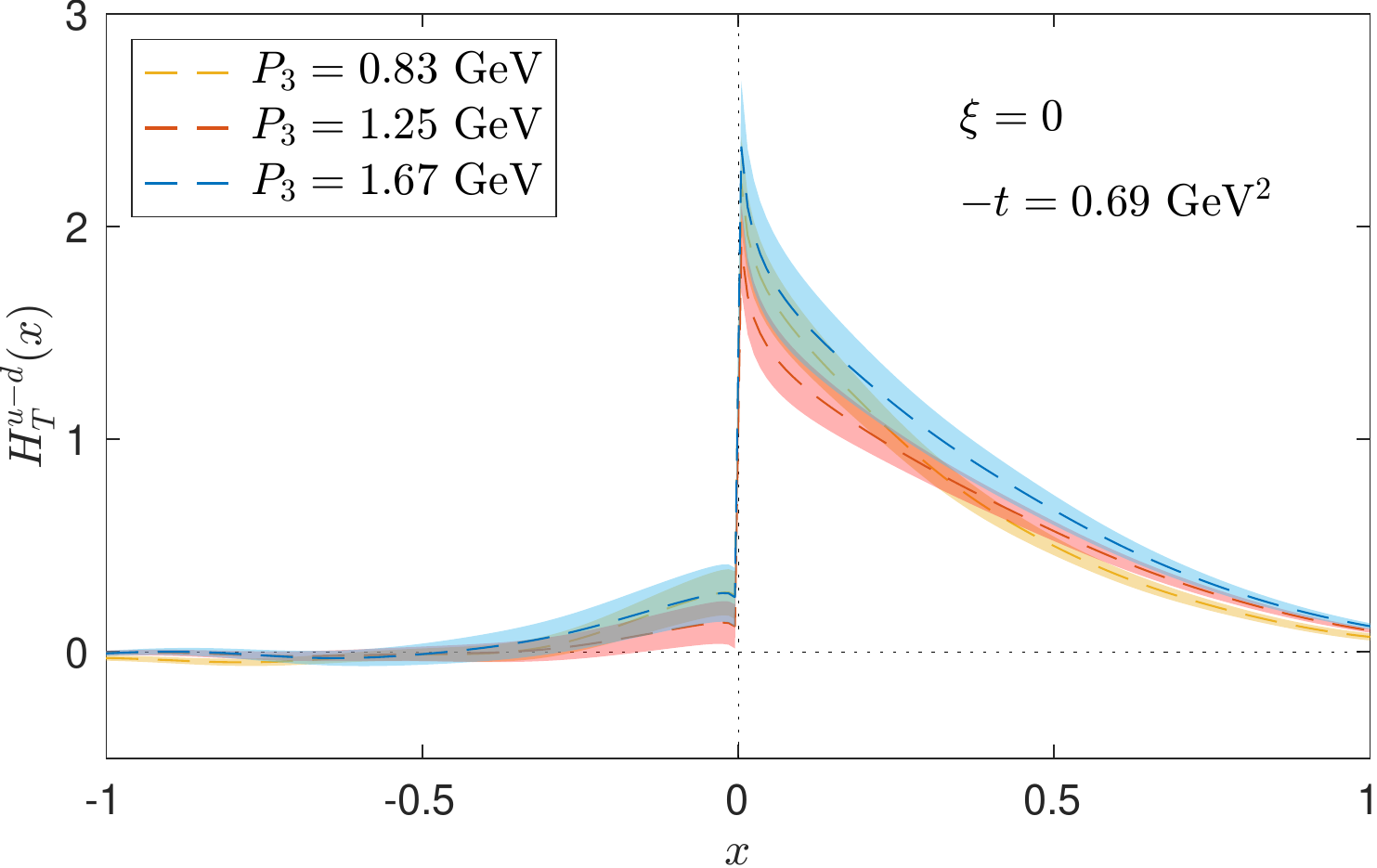}
    \caption{$H$, $E$, $\tilde{H}$ and $H_T$ GPDs at $\xi=0$, $t=-0.69$~GeV$^2$ and for three nucleon boosts: $0.83$~GeV (yellow), $1.25$~GeV (red), $1.67$~GeV (blue). Results are in $\MSb$ scheme at a scale of $2$~GeV.}
    \label{fig:mom_dep_leading}
\end{figure}

\begin{figure}[h!]
    \includegraphics[scale=0.5]{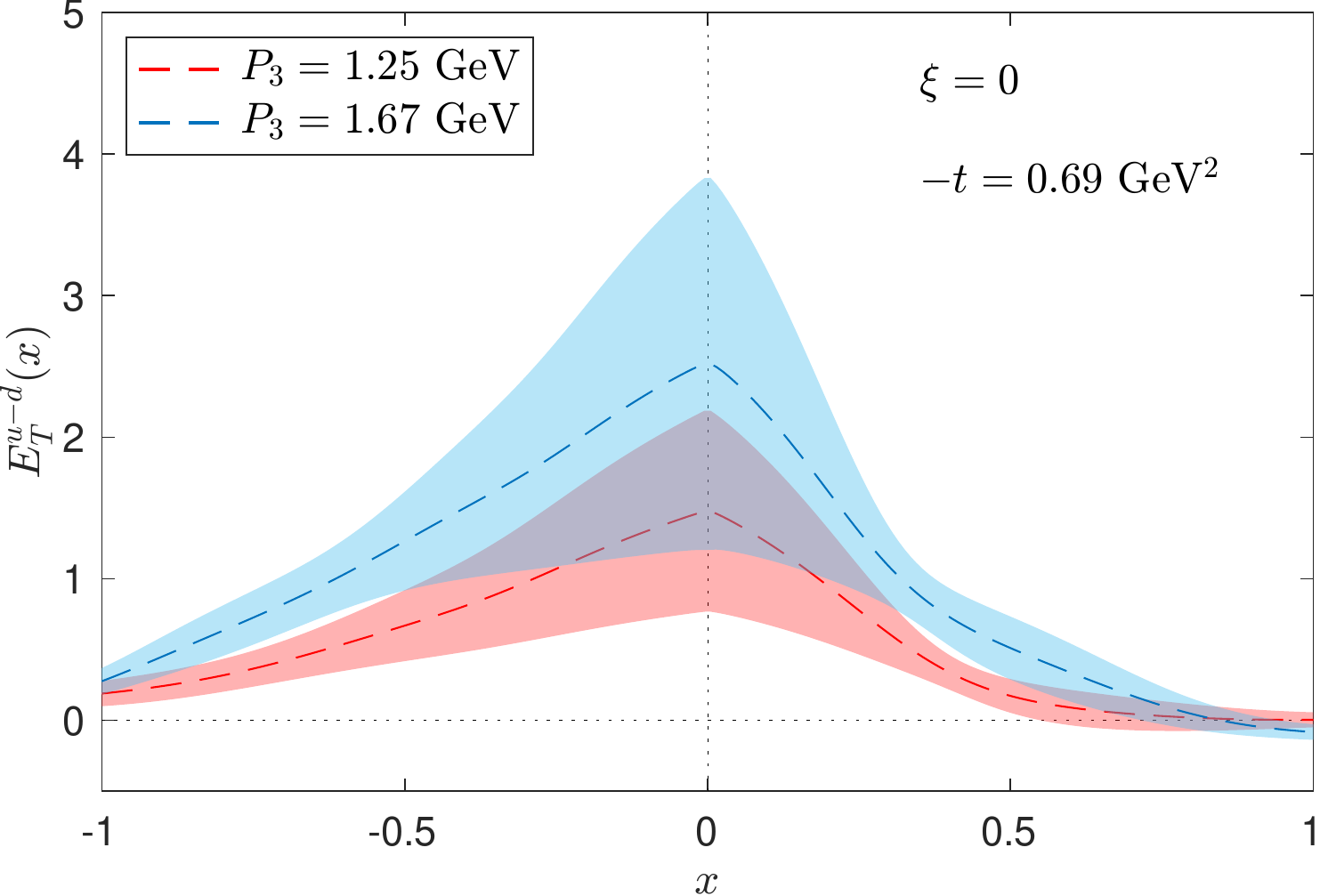}
    \includegraphics[scale=0.5]{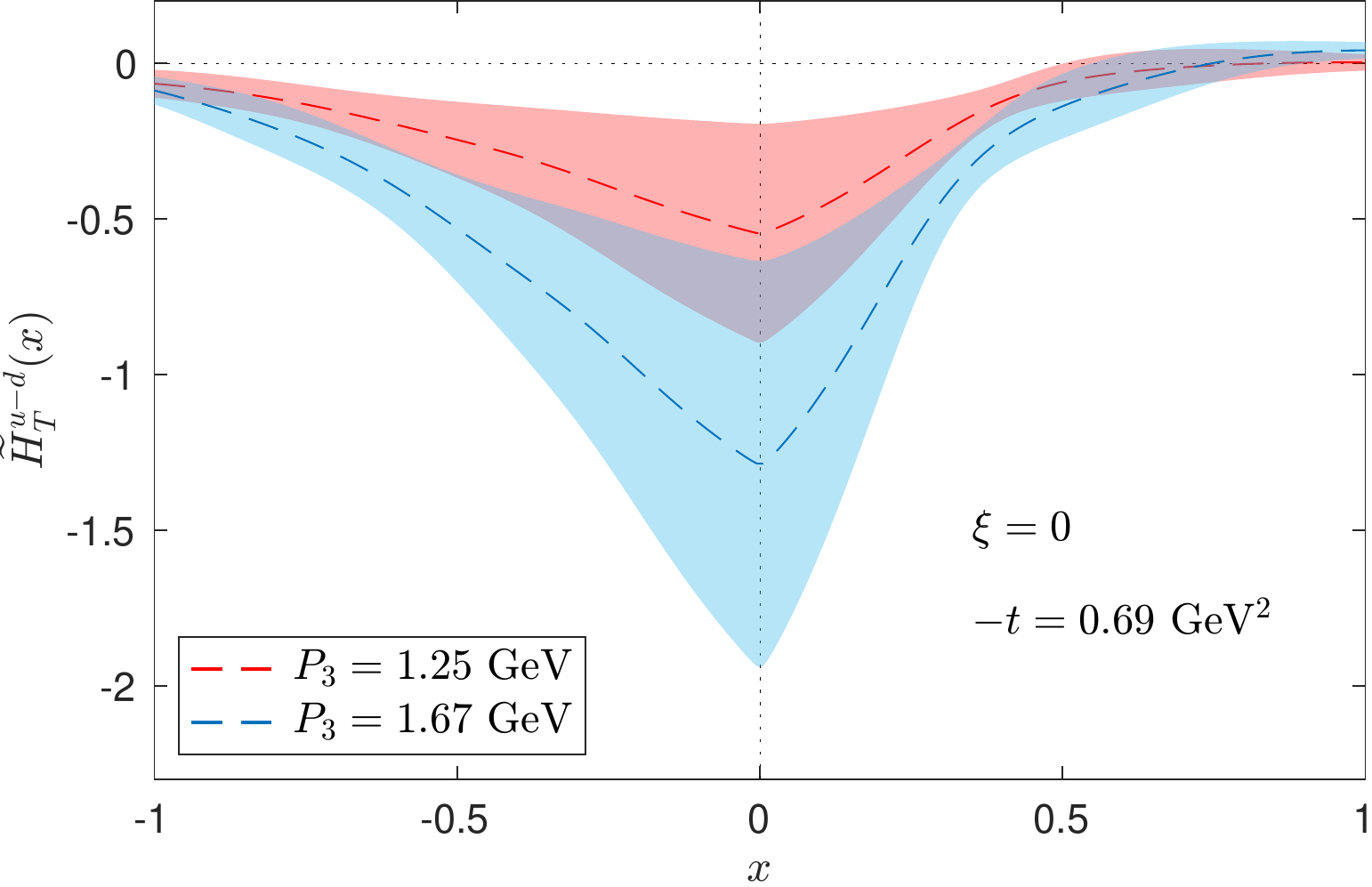}
    \caption{${E}_T$ and $\tilde{H}_T$ at $\xi=0$, $t=-0.69$~GeV$^2$ and proton boosts $1.25$~GeV (red), $1.67$~GeV (blue). Results are in $\MSb$ scheme at a scale of $2$~GeV.}
    \label{fig:mom_dep_subleading}
\end{figure}

Since the whole approach relies on the use of a very large momentum, it is important to check the momentum dependence of the lattice-extracted GPDs. This test is performed at $\xi=0$. The results are reported in Figs.~\ref{fig:mom_dep_leading},\ref{fig:mom_dep_subleading}, where the bands denote the statistical uncertainties. As can be seen, $H$, $\tilde{H}$ and $H_T$ show a very mild $P_3$-dependence and, for the GPDs in Fig.~\ref{fig:mom_dep_leading}, convergence is reached at the two largest boosts. Within larger uncertainties, we also observe compatibility of the results for ${E}_T$ and $\tilde{H}_T$, for which $P_3=0.83$~GeV is not shown since this is a too low momentum. 
We also notice that, at this accuracy, ${E}_T$ and $\tilde{H}_T$ do not show asymmetry between quarks ($x>0$) and antiquarks ($x<0$). We also find that $\tilde{H}_T$ is negative and a similar qualitative conclusion was found within the scalar diquark model~\cite{Bhattacharya:2019cme}.
\subsection{PDFs and GPDs at zero and nonzero skewness}
\begin{figure}[h!]
\begin{minipage}{0.49\textwidth}
    \includegraphics[scale=0.5]{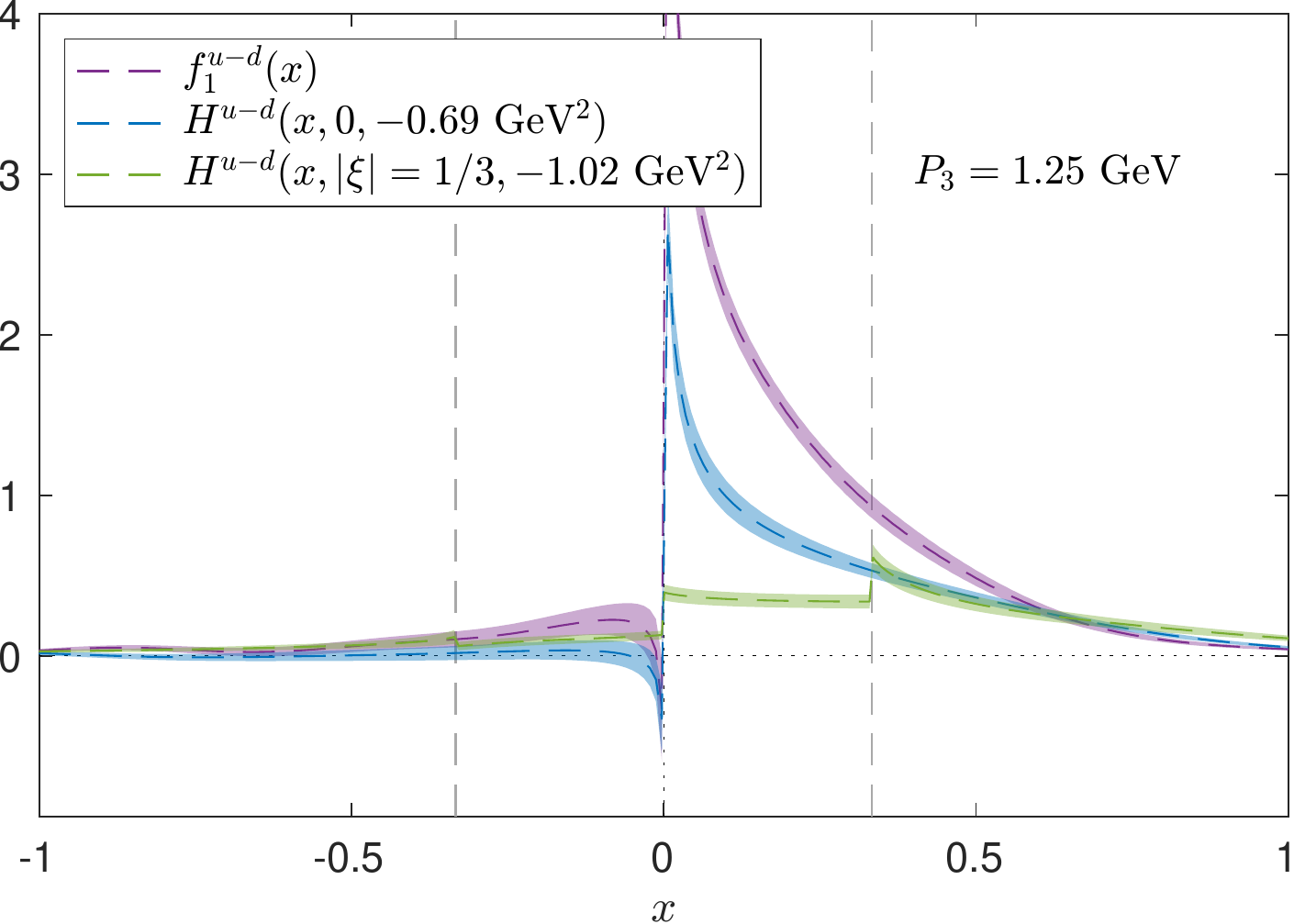}
    \end{minipage}
    \begin{minipage}{0.49\textwidth}
    \includegraphics[scale=0.5]{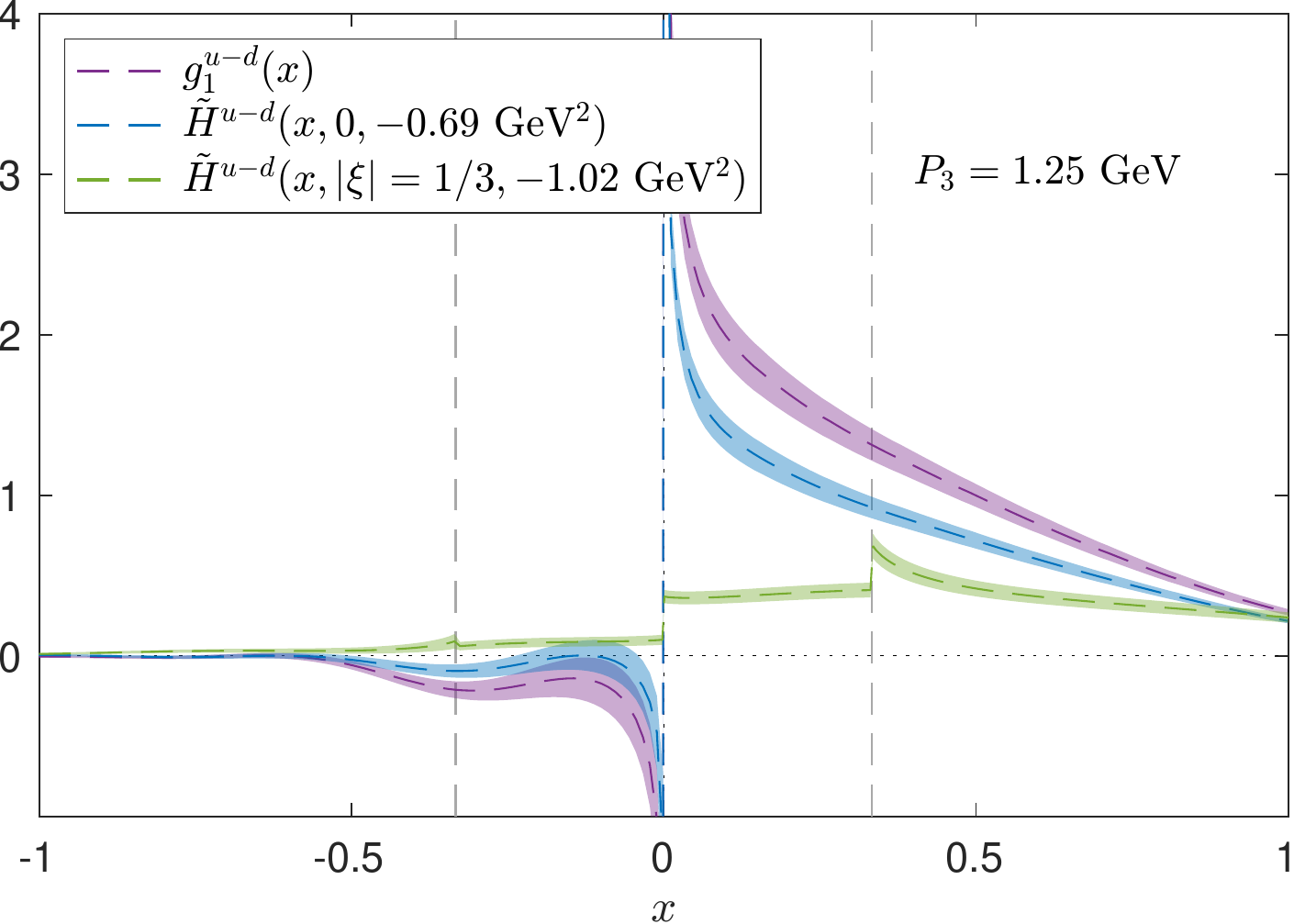}
    \end{minipage}
    
    \begin{minipage}{0.49\textwidth}
    \includegraphics[scale=0.5]{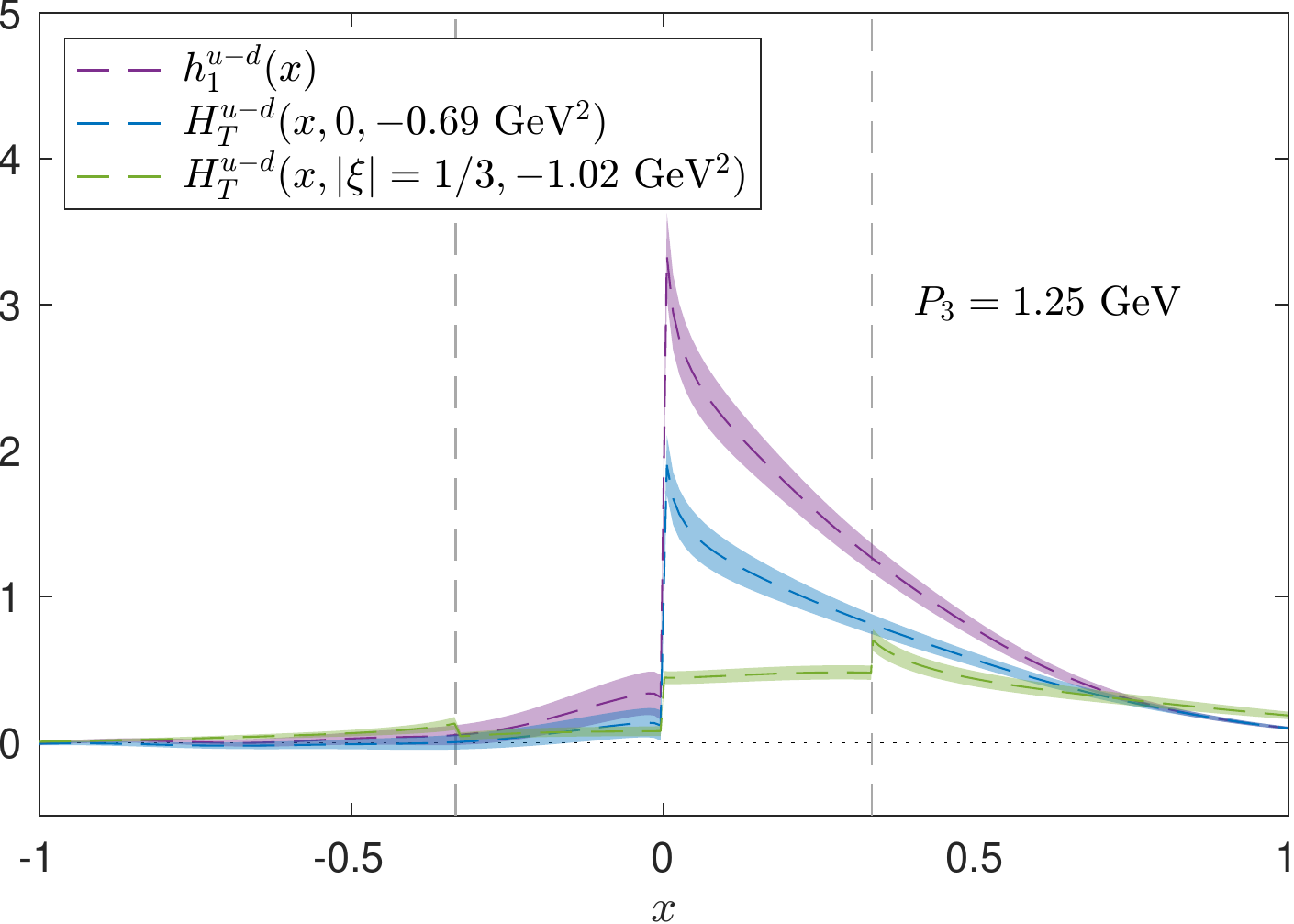}
    \end{minipage}\hspace{0.5cm} 
    \begin{minipage}{0.44\textwidth}
    \caption{Comparison of the PDFs, $f_1(x)$, $g_1(x)$ and $h_1(x)$ (violet band), with the corresponding leading GPDs $H$, $\tilde{H}$ and $H_T$ at $\lbrace \xi=0,-0.69$GeV$^2\rbrace$ (blue band) and $\lbrace  |\xi|=1/3,-1.02$GeV$^2\rbrace$ (green band). The vertical dashed lines delimit the DGLAP ($|x|>1/3$) from the ERBL ($|x|<1/3$) region. Results are given in $\MSb$ at a scale of $2$~GeV.}
    \label{fig:PDFs_GPDs}
    \end{minipage}
\end{figure}

Once momentum convergence of the GPDs at the two largest boosts has been established, we focus on the intermediate momentum, $P_3=1.25$~GeV, and compare the PDFs with the corresponding leading GPDs at $\xi=0$ and $|\xi|=1/3$. The results are shown in Fig.~\ref{fig:PDFs_GPDs}. As we can see, the PDFs are always dominant and the GPDs get suppressed as $-t$ increases, as expected. In fact, a reduction in magnitude with the momentum transfer is also observed in standard form factors, that are just integrals over $x$ of the GPDs.
In addition, at nonzero $\xi$, there is a non-trivial distinction between the  Dokshitzer-Gribov-Lipatov-Altarelli-Parisi (DGLAP; $x>|\xi|$) region and the  Efremov-Radyushkin-Brodsky-Lepage (ERBL; $x<|\xi|$) region, that have a different physical interpretation~\cite{Ji:1998pc} and that are here delimited by dashed vertical lines. For all three cases, we find that the ERBL region is more sensitive to the increase of the momentum transfer and that for $x=\pm \xi$ the results are discontinuous. However, the latter is a non-physical effect  that arises from higher-twist contributions in the matching equations, not yet computed. Despite that, we find agreement between integrals of the GPDs with the corresponding form factors obtained with local operators in the analysis~\cite{Alexandrou:2013joa}. This serves as an important check of the calculation and demonstrates that extracting GPDs within lattice QCD is possible, even though highly non-trivial. It is also worth mentioning that the large-$x$ behavior of $H$ is in qualitative agreement with $1/(1-\xi^2)^2$ behavior predicted by power counting analysis of the unpolarized GPDs~\cite{Yuan:2003fs}. However, further studies, that include assessment of systematics, are necessary for quantitative conclusions.

\section{Conclusions}
In these proceedings, we summarize the first-ever results for unpolarized, helicity and transversity GPDs of the proton. We consider the isovector $u-d$ combination and employ the quasi-distribution formalism, that allows to access GPDs using a finite momentum and LaMET~\cite{Ma:2014jla,Ji:2020ect} to match quasi- to light-cone GPDs. On an $N_f=2+1+1$ ensemble of maximally twisted mass fermions, at pion mass $M_\pi=260$~MeV, we extract the GPDs at $\lbrace \xi=0,-t=0.69$~GeV$^2\rbrace$ and at three values of the boost, and observe that convergence is reached between $P_3=1.25,1.67$~GeV within uncertainties. At $P_3=1.25$~GeV we also test the effect of nonzero skewness and find that ERBL and DGLAP regions are qualitatively different. When comparing GPDs and standard form factors, we see that the GPDs reduce in magnitude as $-t$ increases, and the $n=0$ moment  is compatible with the analysis~\cite{Alexandrou:2013joa} of electromagnetic and axial form factors.

This work will pave the way for more detailed investigations in the future. We plan to study various systematics (related e.g.\ to the lattice spacing, volume effects and pion mass) and also extend the kinematic coverage to multiple values of $\xi$ and $-t$, for which larger volume ensembles will be crucial. In a long-term program, this can give us access to impact parameter distributions and orbital angular momentum of partons, among others. 

\section{Acknowledgements}
The work of K.C. is supported by National Science Centre (Poland) grant SONATA BIS no.~2016/22/E/ST2/00013. M.C. acknowledges financial supported by the U.S. Department of Energy Early Career Award under Grant No.\ DE-SC0020405. K.H. is supported by the Cyprus Research and Innovation Foundation under grant POST-DOC/0718/0100. F.S.\ was funded by DFG project number 392578569. Partial support is provided by the European Joint Doctorate program STIMULATE of the European Union’s Horizon 2020 research and innovation programme under grant agreement No. 765048. Computations for this work were carried out in part on facilities of the USQCD Collaboration, which are funded by the Office of Science of the U.S. Department of Energy. This research was supported in part by PLGrid Infrastructure (Prometheus supercomputer at AGH Cyfronet in Cracow).
Computations were also partially performed at the Poznan Supercomputing and Networking Center (Eagle supercomputer), the Interdisciplinary Centre for Mathematical and Computational Modelling of the Warsaw University (Okeanos supercomputer) and at the Academic Computer Centre in Gda\'nsk (Tryton supercomputer).

\bibliographystyle{JHEP}
{\small
\bibliography{references}}
\end{document}